\documentclass[aps,prl,twocolumn,showpacs]{revtex4}
\usepackage{graphicx}
\newcommand{\ds}{ _{\downarrow}}
\newcommand{\us}{ _{\uparrow}}
\newcommand{\up}{\uparrow}
\newcommand{\down}{\downarrow}

\begin{document}
\draft \title{Normal state of highly polarized Fermi gases:  Full many-body treatment}
\author{R. Combescot and S. Giraud} \address{Laboratoire de
Physique Statistique, Ecole Normale Sup\'erieure, 24 rue Lhomond,
75231 Paris Cedex 05, France}
\date{Received \today}

\begin{abstract}
We consider a single $\down$ atom within a Fermi sea of $\up$ atoms. We elucidate by a full many-body analysis the quite mysterious
agreement between Monte-Carlo results and approximate calculations taking only into account single
particle-hole excitations.
It results from a nearly perfect destructive interference
of the contributions of states with more than one particle-hole pair.
This is linked to the remarkable efficiency of the expansion in powers of hole wavevectors,
the lowest order leading to perfect interference. Going up
to two particle-hole pairs gives an essentially perfect agreement with known exact results.
Hence our treatment amounts to an exact solution of this problem.
\end{abstract}

\maketitle

The last few years have seen a very strong development of works, both experimental
and theoretical, devoted to the exploration of the physical properties of ultracold fermionic
gases \cite{gps}. In particular through the existence of Feshbach resonance these
systems provide a physical realization of the BEC-BCS crossover, and around the resonance
very simple examples of strongly interacting fermionic systems, of high interest for example in
condensed matter physics. More recently much efforts have focused on systems
where there is a possibly strong imbalance between the two fermionic populations
present in these gases \cite{gps, rimit}, corresponding for example to the two lowest energy hyperfine
states of $^6$Li or $^{40}$K.

A very interesting limiting case of such imbalanced mixture is the case of a single fermion
(say $\down$-spin) with mass $m\ds$ in the presence of a Fermi sea of another, say $\up$, 
fermion species with mass $m\us$.
This kind of mixture has anyway to be considered in the analysis \cite{lrgs,fred} of experiments on trapped
gases. Of major importance for the analysis of the phase diagram \cite{lrgs,fred,pg} are the binding energy 
$E_b$ and the effective mass $m^*$ of the $\down$-atom. At unitarity recent fixed-node Monte Carlo (MC)
calculations  \cite{pg} give $(5/3)E_b/E_F =0.99(1)$ where $E_F$ is the Fermi energy of the $\up$ atoms.
On the other hand a simple $T$-matrix analytical calculation, which happens to coincide with a
variational calculation \cite{fred,crlc}, gives $E_b=0.6066 E_F$ which is remarkably close to the MC result.
This closeness has been confirmed very  recently by diagrammatic MC calculations, leading to
$E_b=0.618 E_F$ \cite{ps1} or $E_b=0.615 E_F$ \cite{ps2}, depending on the specific detailed
handling. The very close proximity of the analytical and the MC results, considered to be quite near 
the exact result, is a major puzzle. Indeed at
unitarity the system is strongly interacting, whereas the analytical treatment considers only single
particle-hole excitations (of the non-interacting Hamiltonian $H_c$). 
This  description is only appropriate for a weakly interacting system and 
should fail to a large extent at unitarity.

In this paper we solve this puzzle and at the same time provide an essentially exact solution to the problem.
This is done by considering states with any number of particle-hole excitations. While the ground state
of the interacting Hamiltonian $H$ has important weights coming from many particle-hole excitations,
we show that with respect to the calculation of the energy of the $\down$-atom these states give
contributions which decrease extremely rapidly with the number of particle-hole excitations.
This is basically due to destructive interferences between contributions of these many
particle-hole excitations states when two particles (or holes) are exchanged. The efficiency of
these interferences is directly linked to the key ingredient of our solution, namely the fact that
an expansion in powers of the holes wavevectors ${\bf q}_i$ turns out, quite surprisingly and unexpectedly,
to be an excellent approximation scheme. To lowest order in this approximation, i.e. when the
dependence on the ${\bf q}_i$'s is neglected, the interference is perfect and there is a complete
decoupling between the states with different number of particle-hole excitations. In particular this
justifies the consideration of single particle-hole excitations and explains why the above analytical
result is so good. Improving the approximation by taking more properly into account the ${\bf q}_i$
dependence provides a small coupling between the states with one and two particle-hole excitations, 
providing a small correction to the preceding result. In turn an even better handling of this dependence 
gives a coupling between states with two and three particle-hole excitations, giving a small change to
the preceding small correction, and so on. In this way we obtain a series of results with strongly
increasing accuracy, converging extremely rapidly to the exact result \cite{note}.

Since the theoretical solution of this problem has reached a high and secure level of agreement, 
which is quite unfrequent in many-body problems including the case of cold fermionic gases, 
it provides a useful and convenient benchmark for experiments. Fortunately the agreement with
experiments seems reasonably good \cite{shin}. Moreover our solution is of high general interest
since exact solutions of many-body problems are quite unfrequent, mainly for three dimensional
cases. Our finding that a hole-wavevector expansion works so well is also quite interesting, and
it would be worthwhile to apply it to more complex many-body problems. Another useful conclusion 
regarding self-consistent treatments can also be taken out of our solution. An attractive and standard
way to try to improve on a simple $T$-matrix approximation, such as the one used in \cite{crlc},
is to use full propagators rather than bare ones in the $T$-matrix. The practical handling is
much more complicated, but this seems much more satisfactory since this amounts to include an
infinite set of higher order diagrams. This has been done for example recently in the BEC-BCS crossover
for equal spin populations \cite{hauss}. We have made the corresponding self-consistent treatment
for the $T$-matrix approximation of Ref.\cite{crlc}. The result at unitarity is $E_b=0.68\,E_F$, much 
worst than the non self-consistent one. Our solution allows to understand this failure. Indeed the exact
solution displays almost complete cancellations between diagrams, whereas the self-consistent treatment destroys this delicate balance since it retains only some diagrams at a given order. Hence
self-consistency does not systematically lead to an improvement.

Here we concentrate on the unitarity case. Indeed in the weak limit $a\rightarrow 0_{-}$,
an expansion in powers of $a$ should be valid and the lowest order correction should
give already quite good results. On the other side of unitarity, one can easily see from 
the details given below that our approximation should improve for increasing $1/a>0$.
Hence unitarity is a kind of worst case for our approach. Since it attracts in addition
most of the interest in the strong coupling regime of large $a$, this justifies that we
restrict our discussion to this case. Generalization away from unitarity is quite obvious.

Before presenting our solution, we display very explicitely the efficiency of the ${\bf q}_i$ expansion
on the simplest case of the lowest order solution \cite{crlc}. At unitarity, for equal masses $m\us=m\ds=m$,
the binding energy $E_b=-E=|E|$ is given by:
\begin{eqnarray}\label{eqlow}
\hspace{-17mm}|E|=\sum_{q<k_F}
\left[\sum_{k}\!\frac{m}{k^{2}}-\sum_{k>k_F} \frac{1}
{E^{(1)}_{\bf kq}}
\right]^{-1}
\end{eqnarray}
where $E^{(1)}_{\bf kq}= |E|+\epsilon_{{\bf k}-{\bf q}} + \epsilon _{\bf k}- \epsilon _{\bf q}$
with $\epsilon_{\bf k} = {\bf k}^{2}/2m$. Setting ${\bf q}={\bf 0}$ in $E^{(1)}_{\bf kq}$ leads to the equation
$2/3\rho=1+\sqrt{\rho/2}\arctan{\sqrt{\rho/2}}$ for $\rho=|E|/E_F$. The solution $\rho=0.5347$
is not so far from the actual result, which is already pretty satisfactory for such a crude approximation.
However we may quite improve on it by treating ${\bf q}$ to lowest significant order instead 
of neglecting it completely. This means we take for the ${\bf k}$ angular average
$\langle\left[|E|+\!\epsilon _{{\bf k}-{\bf q}}+\epsilon_{{\bf k}}-\!\epsilon_{{\bf q}}\right]^{-1}\rangle_{\bf k}
\simeq \left[2\epsilon_{k}+|E|\right]^{-1}+q^2k^2/(3m^2)\left[2\epsilon_{k}+|E|\right]^{-3}$. This gives at unitarity $\rho=0.5985$, which is remarkable compared to 
the exact \cite{crlc} solution $\rho=0.6066$ of Eq.(\ref{eqlow}) and the MC result $(5/3)\rho =0.99(1)$ \cite{pg}. 
If we do not restrict ourselves to unitarity and let the scattering length vary, the corresponding solution 
for $\rho$ turns out to be essentially undistinguishable on a graph from the exact numerical solution \cite{crlc}. Although the situation is somewhat less favourable for unequal masses, this shows very simply and explicitely the validity and efficiency of this ${\bf q}_i$ expansion approach. We apply it now to the full many-body problem.

The Hamiltonian of our problem is:
\begin{eqnarray}\label{}
H&=&H_c+V \\
H_c&=& \sum_{\bf P}E({\bf P})b^{\dag}_{{\bf P}}b_{\bf P}+ \sum_{\bf k}\epsilon _{\bf k}c^{\dag}_{\bf k}c_{\bf k}\\
V&=&g \sum_{\bf k k' P P'}\delta_{\bf k k' P P'}c^{\dag}_{\bf k}c_{\bf k'}b^{\dag}_{{\bf P}}b_{\bf P'}
\end{eqnarray}
where $\epsilon_{\bf k} ={\bf k}^{2}/2m_{\uparrow}$, $\,E({\bf P})={\bf P}^2/2m\ds$, 
and $c_{{\bf k}}$ and $c^{\dag}_{{\bf k}}$ are
annihilation and creation operators for $\uparrow$-spin atoms while $b_{{\bf P}}$ and $b^{\dag}_{\bf P}$ are for the $\downarrow$-spin atom. In the potential energy term $V$, the short-ranged
interaction potential provides an upper cut-off $k_c$, which we will as usual let go to infinity while the coupling constant $g$ goes to zero, keeping the scattering length finite in the relation $m_r/(2\pi a)=g^{-1}+ \sum_{}^{k_c}2m_r/ k^2$,
where $m_r = m\us m\ds /(m\us + m\ds)$ is the reduced
mass. Momentum conservation in the scattering is insured by the
Kronecker symbol $\delta_{\bf k k' P P'}$. Physically this potential energy term creates (or annihilates) a single particle-hole pair from the Fermi sea, or merely scatters particles or holes, the momentum transfer
being taken by the $\downarrow$-spin atom. Hence we are looking for the ground state as a general superposition of states with any number of particle-hole pairs:
\begin{eqnarray}\label{}
|\psi\rangle&=&\alpha _0 |0\rangle+ \sum_{\bf kq}\alpha _{\bf kq}c^{\dag}_{\bf k}c_{\bf q}|0\rangle+\cdots \\\nonumber
&&+\frac{1}{(n!)^2}\sum_{\{{\bf k}_i\}\{{\bf q}_j\}}\alpha _{{\bf k}_i{\bf q}_j}\prod_{i=1}^{n} c^{\dag}_{{\bf k}_i}\prod_{j=1}^{n} c_{{\bf q}_j}|0\rangle+\cdots
\end{eqnarray}
where $|0\rangle=\prod_{k<k_F} c^{\dag}_{\bf k}\,|vac\rangle$ is the Fermi sea of $\uparrow$-spins. For simplicity, in all the terms, we have not written explicitely the creation operator $b^{\dag}_{{\bf P}}$ of the $\downarrow$-spin since its momentum 
${\bf P}$ is just obtained to insure that the total momentum of the particles in any term is zero. We assume also implicitely $k_i>k_F$ and $q_j<k_F$. The coefficients
$\alpha _{{\bf k}_i{\bf q}_j}$ (which is a shorthand for $\alpha _{\{{\bf k}_i\}\{{\bf q}_j\}}$) are naturally antisymmetric with respect to the exchange of any their arguments ${\bf k}_i$ or ${\bf q}_j$.

Writing $H|\psi\rangle=E|\psi\rangle$ (where we take as zero for
the energy $E$ the energy of the whole Fermi sea, and omit the average potential energy
which disappears for $g\rightarrow0$) 
and identifying the coefficients of specific particle-hole states in both terms, we find
for the equations corresponding to the full Fermi sea and the Fermi sea with a single particle-hole pair:
\begin{eqnarray}\label{eq0}
\hspace{-30mm}-g^{-1}|E|\alpha _0\!=\!\sum_{\bf kq}\alpha _{\bf kq}\hspace{14mm}
\end{eqnarray}
\vspace{-5mm}
\begin{eqnarray}\label{eq1}
\!\!-g^{-1}E^{(1)}_{{\bf kq}}\alpha _{\bf kq}\!=\!\alpha _0+ \!\sum_{{\bf K}} \alpha _{\bf Kq}-
\!\sum_{{\bf Q}} \alpha _{\bf kQ}- \sum_{{\bf K}{\bf Q}}\alpha _{{\bf k}{\bf K}{\bf q}{\bf Q}}
\end{eqnarray}
where we have introduced $E^{(n)}_{{\bf k}_i{\bf q}_j}=|E|+E(\sum_i^n{{\bf k}_i}- \sum_j^n{{\bf q}_j})
+ \sum_i^n\epsilon _{{\bf k}_i} -  \sum_j^n\epsilon _{{\bf q}_j}$. In the second equation the first term comes from the creation of a particle-hole pair in the Fermi sea, the second from the diffusion of the particle, the third from the diffusion of the hole and the last one from the annihilation of a particle-hole pair. In the last term, from two particle-hole pair states, there are four combinations for annihilation of a particle-hole pair (this would be $n^2$ in the general case). The next equation from two particle-hole pairs states is:
\begin{eqnarray}\label{eq2}
-g^{-1}E^{(2)}_{\bf kk'qq'}\alpha _{\bf kk'qq'}\!=\!-\alpha _{\bf kq}\!-\alpha _{\bf k'q'}\!+\alpha _{\bf kq'}\!+\alpha _{\bf k'q} \hspace{5mm} \nonumber
\end{eqnarray}
\vspace{-5mm}
\begin{eqnarray}
+ \!\sum_{{\bf K}} \alpha _{\bf Kk'qq'}\!+\!\sum_{{\bf K}} \alpha _{\bf kKqq'}\!-\!\!\sum_{{\bf Q}} \alpha _{\bf kk'Qq'}\!-\!\!\sum_{{\bf Q}} \alpha _{\bf kk'qQ} \hspace{3mm}\nonumber
\end{eqnarray}
\vspace{-5mm}
\begin{eqnarray}\label{eq2}
+\sum_{{\bf K}{\bf Q}}\alpha _{{\bf Kkk'}{\bf Qqq'}} \hspace{55mm}
\end{eqnarray}
with physical origins for the various terms analogous to the preceding ones. One could write formally the generalization for any order, but we will not do it since this will not be necessary.

In Eq.(\ref{eq1}) the second term turns out to give a divergent contribution when $k_c\rightarrow\infty$
and the limit is finite only after multiplication by $g$. By comparison the third term, where the summation
is on the Fermi sea, displays no divergence and gives a zero contribution after multiplication by $g$.
Hence we omit it from now on. Similarly  in Eq.(\ref{eq2}) we can omit the two analogous sums over
${\bf Q}$.

%We may now argue as above in the limit of large $|E|$ to show that in this case we may take 
%$E^{(n)}_{{\bf k}_i{\bf q}_j} \simeq |E|+E(\sum_i^n{{\bf k}_i}) + \sum_i^n\epsilon _{{\bf k}_i} \equiv
%E^{(n)}_{{\bf k}_i}$. For example in Eq.(\ref{eq2}), we may take $E^{(2)}_{\bf kk'qq'}
% \simeq |E|+E({\bf k}+{\bf k}') + \epsilon _{\bf k}+\epsilon _{\bf k}'=E^{(2)}_{{\bf k}{\bf k}'}$. We can now
% sum Eq.(\ref{eq2}) over ${\bf q}'$. Because of the antisymmetry of $\alpha _{{\bf Kkk'}{\bf Qqq'}}$ 
% in the exchange of ${\bf Q}$ and ${\bf q}'$, the last term drops out. Hence, through this summation, Eq.(\ref{eq2})
% provides us with an integral equation for the function $\sum_{{\bf K}{\bf Q}}\alpha _{{\bf k}{\bf K}{\bf q}{\bf Q}}$, 
% which is the only quantity entering Eq.(\ref{eq1}) from the subspace of two particle-hole pairs added to
% the Fermi sea. We see that our approximation make the subspaces with higher number of particle-hole pairs
% irrelevant for our problem, and we are completely decoupled from these higher subspaces. We have obtained, together with Eq.(\ref{eq0}) and Eq.(\ref{eq1}) a closed set of equations for $\alpha _0,\alpha _{\bf kq}$ and
% $\sum_{{\bf Q}}\alpha _{{\bf k}{\bf K}{\bf q}{\bf Q}}$, which can clearly be solved exactly by numerical means
% to provide the energy $|E|$.
 
We see that, to solve our problem, the
only thing we need to know from Eq.(\ref{eq2}) is $\sum_{{\bf K}{\bf Q}}\alpha _{{\bf k}{\bf K}{\bf q}{\bf Q}}$,
which is the only quantity entering Eq.(\ref{eq1}) from the subspace of two particle-hole pairs added to
the Fermi sea.  
Now, if we neglect the wavevectors ${\bf q}$ and ${\bf q}'$ in $E^{(2)}$
and take $E^{(2)}_{\bf kk'qq'} \simeq E^{(2)}_{{\bf k}{\bf k}'{\bf 0}{\bf 0}}$, we can
 sum Eq.(\ref{eq2}) over ${\bf q}'$. Because of the antisymmetry of $\alpha _{{\bf Kkk'}{\bf Qqq'}}$ 
 in the exchange of ${\bf Q}$ and ${\bf q}'$, the last term drops out. Hence, through this summation, Eq.(\ref{eq2})
 provides us with an integral equation for the function $\sum_{{\bf K}{\bf Q}}\alpha _{{\bf k}{\bf K}{\bf q}{\bf Q}}$. 
 We see that our approximation makes the subspaces with higher number of particle-hole pairs
 irrelevant for our problem, and we are completely decoupled from these higher subspaces. We have obtained, together with Eq.(\ref{eq0}) and Eq.(\ref{eq1}) a closed set of equations for $\alpha _0,\alpha _{\bf kq}$ and
 $\sum_{{\bf K}{\bf Q}}\alpha _{{\bf k}{\bf K}{\bf q}{\bf Q}}$, which can clearly be solved exactly by numerical means
 to provide the energy $|E|$. The precise justification for our approximation is that $k , k' > k_F$
 and $q, q'<k_F$ make indeed in $E^{(2)}_{\bf kk'qq'}$ the $k$ and $k' $ terms dominant and $q$ and $q'$ ones 
 negligible in a large part of the variables space. Naturally this is an approximation, the interference is not
 exact but not so far from it. Hence the contribution from three particle-hole pairs states is not zero, but it is quite
 small which is the essential point.
 
%Introducing $G({\bf k},{\bf q},{\bf q}')= g \sum_{{\bf K}}\alpha _{{\bf k}{\bf K}{\bf q}{\bf q}'}$ we see that we
%only need in Eq.(\ref{eq1}) to know $ \sum_{{\bf Q}}G({\bf k},{\bf q},{\bf Q})$ from Eq.(\ref{eq2}) to solve our problem.  

%In turn
%$G({\bf k},{\bf q},{\bf q}')$ is obtained by dividing Eq.(\ref{eq2}) by $E^{(2)}_{\bf kk'qq'}$
%and summing over ${\bf k}'$.
  
 The same argument applied at the level of Eq.(\ref{eq1}) allows now to understand the success of the lowest
 order approach. If we make the approximation $E^{(1)}_{\bf kq}
 \simeq |E|+E({\bf k}) + \epsilon _{\bf k}$ and sum Eq.(\ref{eq1}) over ${\bf q}$, we see that the
 last term is disappearing, due to the antisymmetry of $\alpha _{{\bf k}{\bf K}{\bf q}{\bf Q}}$ with respect to the
 exchange of ${\bf q}$ and ${\bf Q}$, which leaves us with a decoupled set of equations for
 $\alpha _0$ and $ \sum_{{\bf k}{\bf q}}\alpha _{\bf kq}$, leading to the good approximation discussed at the beginning.
 Conversely we may decide to make the approximation $E^{(n)}_{{\bf k}_i{\bf q}_j} \simeq E^{(n)}_{{\bf k}_i}$ at
 some higher order $n$. This provides by the same procedure a decoupling from higher order subspaces
 resulting in a closed  set of $n$ equations to be solved for $|E|$.
 Hence our analysis leads us to a cascade of successive approximations which converges very rapidly in a
 controlled way to the exact many-body solution, providing at the same time a full understanding of the success of the lowest order approximations and a practical way to obtain the exact results within the precision we like.
 
 In practice the convergence is so fast, as shown by the quality of the lowest order results, that we will only need for any practical purpose to implement it to the next (i.e. second) order. However, in addition to our main scheme of approximations going
 from one order to the next, we may within a given order introduce an additional graduation of approximations. First we
 can as described above make the approximation 
 $E^{(2)}_{\bf kk'qq'} \simeq E^{(2)}_{\bf k k' 0 0}$ and sum Eq.(\ref{eq2}) over ${\bf q}'$. However we can easily
 improve on this handling in the same spirit as for the lowest order approximation. Indeed the argument is still valid
 if we make only the approximation $E^{(2)}_{\bf kk'qq'} \simeq E^{(2)}_{{\bf k}{\bf k}'{\bf q}{\bf 0}}$, which takes better into
 account the ${\bf q}$ dependence. Finally since we know that the last term in Eq.(\ref{eq2}) brings only a small correction,
 we may just decide to omit it, but treat the rest of the equation exactly without any approximation on $E^{(2)}_{\bf kk'qq'}$,
 which takes even better account of the ${\bf q}_i$ dependence. This handling is at the same level as treating Eq.(\ref{eqlow}) exactly. In the same way it corresponds to a variational
 calculation since we have restricted the Hilbert space to contain at most two particle-hole pairs. This is the most precise
 level of approximation we will implement.
  
Introducing $G({\bf k},{\bf q},{\bf q}')= g \sum_{{\bf K}}\alpha _{{\bf k}{\bf K}{\bf q}{\bf q}'}$, dividing Eq.(\ref{eq2})
by $E^{(2)}_{{\bf k k' q q'}}$, summing over ${\bf k}'$, taking the $k_c\rightarrow\infty,g\rightarrow 0$ limit,
which makes again some terms vanish, we find the integral equation:
% Introducing $F({\bf k},{\bf q})=
% g \sum_{{\bf K}{\bf Q}}\alpha _{{\bf k}{\bf K}{\bf q}{\bf Q}}$, we obtain an equation for this function by dividing Eq.(\ref{eq2})
% by $E^{(2)}_{{\bf k}{\bf k}'}$ and summing over ${\bf k}'$. Taking the $k_c\rightarrow\infty,g\rightarrow 0$ limit,
% which makes again some terms vanish, we find:
% \vspace{-3mm}
\begin{eqnarray}\label{eq2var}
G({\bf k},{\bf q},{\bf q}')\left[ \frac{m_r}{2\pi a}+\sum_{\bf k'}\left(\frac{1}{E^{(2)}_{\bf kk'qq'}}-\frac{2m_r}{k'^2}\right)\right] 
\hspace{10mm} \nonumber
\end{eqnarray}
\vspace{-5mm}
\begin{eqnarray}
\hspace{15mm}=\alpha _{\bf kq'}-\alpha _{\bf kq}+ \sum_{\bf k'}\frac{G({\bf k}',{\bf q},{\bf q}')}{E^{(2)}_{\bf kk'qq'}}
\end{eqnarray}
The bracket in the left-hand side can be calculated analytically. In practice this integral equation is easily solved
by iteration, the already fast convergence being increased by acceleration procedures. The same handling allows
to solve the set of Eq.(\ref{eq1},\ref{eq2var}) with respect to $\alpha _{\bf kq}$.

\begin{figure}
%\vspace{-2mm}
%\begin{figure}[htbp]
%\begin{center}
\hspace{-10mm}
%\scalebox{0.65}{\rotatebox{270}{\includegraphics[width=10cm]{fig1tot}}}
%\scalebox{0.65}{{\includegraphics[width=10cm]{fig1}}}
\scalebox{0.65}{{\includegraphics[width=10cm]{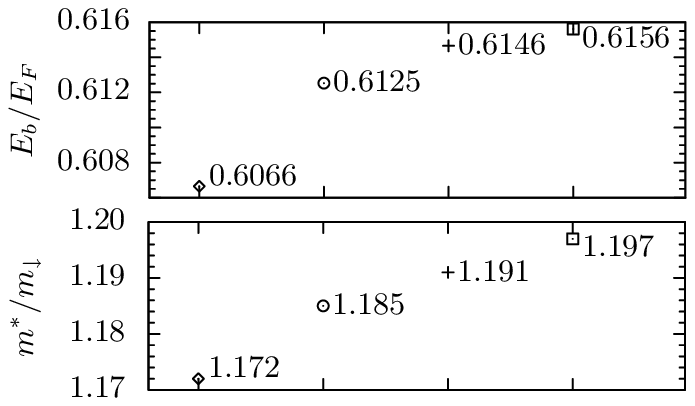}}}
\caption{Reduced $\down$ atom binding energy $E_b$ and effective mass  $m^*$, in the case of
$\up$ and $\down$ atoms with equal masses, for the approximations
of increasing accuracy discussed in the text. $E_F$ is the Fermi energy of the $\up$ atoms.
Diamond: first order approximation \cite{crlc}.
Circle: second order with ${\bf qq'}={\bf 00}$. Cross: second order with ${\bf qq'}={\bf q0}$.
Square: second order with no ${\bf q}$ approximation (variational). For the effective mass we give
coherently the last digits of our results to display clearly the trend, but they should not be taken for granted.}
\label{fig1}
%\end{center}
\vspace{-6mm}
\end{figure}

We have first checked the convergence of our theoretical scheme in the particular case $m\ds\rightarrow \infty$ where
the binding energy is known \cite{crlc} to be exactly $\rho=1/2$ at unitarity. This is naturally a very convenient case. However
one can easily see that the situation, with respect to the convergence of the ${\bf q}_i$ expansion, is slightly less
favorable than for the equal mass case. This is appearant in the first order result \cite{crlc} $\rho=0.465$ which, while being fairly
good, is not so close to $1/2$. A convenient feature of this case is that there are no angular integrations to perform in
solving the various equations. At second order we find 0.481 when we take ${\bf qq'}={\bf 00}$. This improves into
0.487 for ${\bf qq'}={\bf q0}$. However with no approximation on $E^{(2)}$ (i.e. the variational result) we find 
$\rho=0.498$, which shows quite explicitely that essentially, at second order, we have already fully converged toward the
exact result. The relative sensitivity in this case to the approximations on ${\bf q}$ may be physically understood as related to the
lack of quantum smearing from the $\down$-atom since it acts as a fixed impurity. Another situation where the exact solution
is known is the 1D case \cite{crlc, McGuire}. Here again we have checked \cite{1D} that the second order coincides almost with the exact results,
including for the calculation of the effective mass \cite{McGuire}.

We have then proceeded to the same calculations for the equal mass situation (see Fig.\ref{fig1}), which is markedly better in terms
of convergence than the infinite mass case. For the ${\bf qq'}={\bf 00}$ approximation
we find 0.6125, the ${\bf qq'}={\bf q0}$ result is 0.6146. Finally the variational result is 0.6156. This displays quite clearly
the very fast convergence of the ${\bf q}_i$ expansion. It is very likely that this last result is essentially exact \cite{note1} 
(say, the exact
one is bounded by 0.6158). This is supported by how close our above result for infinite $m\ds$ is from the exact one. More
precisely, as seen explicitely from the $m\ds=\infty$ case,
each order brings a small correction to the preceding one. Since the second order result brings typically a $10^{-2}$
correction to the first order one, we expect the third order correction to be at most of order $10^{-4}$. Naturally the precision
of our result is beyond practical use. We display it only to support our claim that we have a full solution for our problem.

Finally we have calculated the effective mass, with the same approach applied to a small nonzero momentum for
the whole system. This produces only some minor practical complications. We find $m^*/m\ds=1.20$. Our above
precision on the binding energy gives us typically a $\pm 0.02$ uncertainty. In agreement with our above findings, 
this result is fairly close to the first order value 1.17. We see that there is some significant discrepancies with some 
MC results \cite{lrgs,pg} as well as conclusions from experiments \cite{shin}. Though they are not major ones, they
may be relevant for the detailed understanding of experiments \cite{rls}.

% and are clearly
%linked the effective mass is more sensitive than the binding energy.

We are grateful to X. Leyronas for stimulating discussions. The ``Laboratoire de
Physique Statistique'' is ``Laboratoire associ\'e au Centre National de la Recherche
Scientifique et aux Universit\'es Paris 6 et Paris 7''.


\begin{references}
\bibitem{gps}For a very recent review, see S. Giorgini, L. P. Pitaevskii and S. Stringari, arXiv:0706.3360
and to be published in Rev. Mod. Phys.
\bibitem{rimit}G. B. Partridge, W. Li, R. I. Kamar, Y. Liao and
R. G. Hulet, Science {\bf 311}, 503 (2006);Y. Shin, M. W. Zwierlein, C. H. Schunck, A. Schirotzek,
and W. Ketterle, Phys. Rev. Lett. {\bf 97}, 030401
(2006); C. H. Schunk, Y. Shin, A.Schirotzek, M. W. Zwierlein, and
W. Ketterle, Science {\bf 316}, 867 (2007); Y. Shin, C. H. Schunck, A. Schirotzek, and W. Ketterle,
Nature {\bf 451}, 689 (2008).
\bibitem{lrgs}C. Lobo, A. Recati, S. Giorgini and S. Stringari,
Phys. Rev. Lett. {\bf 97}, 200403 (2006).
\bibitem{fred}F. Chevy, Phys. Rev. A {\bf 74}, 063628 (2006).
\bibitem{pg}S. Pilati and S. Giorgini, Phys. Rev. Lett. {\bf 100}, 030401 (2008).
\bibitem{crlc}R. Combescot, A. Recati, C. Lobo and F. Chevy,
Phys. Rev. Lett. {\bf 98}, 180402 (2007).
\bibitem{ps1}N. V. Prokof'ev and B. V. Svistunov, Phys. Rev. B {\bf 77}, 020408 (2008).
\bibitem{ps2}N. V. Prokof'ev and B. V. Svistunov, Phys. Rev. B {\bf 77}, 125101 (2008).
\bibitem{note}Note that the existence of strong cancellations between diagrammatic contributions
has been pointed out in Ref. \cite{ps1} and \cite{ps2} in the course of MC calculations. Our solution
provides an organization of the various terms, allowing to understand and to make use of the
nearly complete cancellations.
\bibitem{shin}Y. Shin, Phys. Rev. A {\bf 77}, 041603(R) (2008) and references therein
\bibitem{hauss}R. Haussmann, W. Rantner, S. Cerrito, and W. Zwerger, Phys. Rev. A {\bf 75}, 023610 (2007)
\bibitem{1D}S. Giraud and R. Combescot, in preparation
\bibitem{note1}It agrees very nicely with the MC results of \cite{ps2}
\bibitem{McGuire} J. B. McGuire, J. Math. Phys. (N.Y.) {\bf 7}, 123 (1966).
\bibitem{rls}A. Recati, C. Lobo and S. Stringari, arXiv:0803.4419
\end{references}
\end{document}